\documentclass[11pt]{article}
\usepackage{graphicx}
\usepackage{amsmath}
\usepackage{amsfonts}
\usepackage{amssymb}
\usepackage{MnSymbol}

\numberwithin{equation}{section}

\newcommand{\be}{\begin{equation}}
\newcommand{\ee}{\end{equation}}

\newcommand{\M}{\mathcal{M}}

\begin{document}

\title{Elementary derivation of Weingarten functions of classical Lie groups}
\author{Marcel Novaes\\
{\small Instituto de F\'isica, Universidade Federal de Uberl\^andia, Uberl\^andia, MG, 38408-100, Brazil}}\date{} \maketitle

\begin{abstract}
Integration of polynomials over the classical groups of unitary, orthogonal and
symplectic matrices can be reduced to basic building blocks known as Weingarten
functions. We present an elementary derivation of these functions.
\end{abstract}
%\maketitle
\section{Introduction}

Integration with respect to matrix ensembles is the core of random matrix theory
\cite{Mehta} and an important problem in many areas of mathematical physics.
\cite{handbook} Classical Lie groups of unitary, orthogonal and symplectic matrices,
endowed with the corresponding Haar measure, constitute important classes of matrix
ensembles. Integrals of functions which are polynomials in the matrix elements can be
reduced to the general form \be \int_G \prod_{k=1}^{n_1} U_{i_kj_k} \prod_{\ell=1}^{n_2}
U^*_{m_\ell r_\ell} dU\equiv\left\langle\prod_{k=1}^n U_{i_kj_k}\prod_{\ell=1}^{n_2}
U^*_{m_\ell r_\ell}\right\rangle_G,\ee where $G$ denotes one of the groups and $dU$ the
corresponding Haar measure.

Initial investigations in the physics literature considered the unitary group
$\mathcal{U}(N)$.\cite{wein,creutz,samuel,mello,BB,aubert,esposti} It was found that the
basic integral can be written as a double sum over the symmetric group, \be
\left\langle\prod_{k=1}^n U_{i_kj_k}
U^*_{m_kr_k}\right\rangle_{\mathcal{U}(N)}=\sum_{\tau,\sigma\in S_n} {\rm
Wg}^U(\tau^{-1}\sigma)\prod_{k=1}^n[i_k=m_{\sigma(k)}][j_k=r_{\tau(k)}],\ee where $[i=m]$
is the same as $\delta_{i,m}$ and Wg$^U$ is a function which depends only on the cycle
structure of its argument. Independently of the physicists, Collins \cite{collins}
rediscovered the problem and suggested Wg$^U$ be called the Weingarten function of
$\mathcal{U}(N)$.

The corresponding question for the orthogonal and symplectic groups has also been much
studied. \cite{gorin,prosen,lopez,braun} They too can be written as double sums over
permutations, and the corresponding Weingarten functions Wg$^O$ and Wg$^{Sp}$ have been
found. \cite{CS,CM} Later developments include new approaches (e.g. from Jucys-Murphy
elements, \cite{NovakJucys,zinn}) generalizations (e.g. to compact symmetric spaces
\cite{stolz,mats}) connections (e.g. to factorizations of permutations \cite{MN}) and
applications (e.g. to different polynomial integrals, \cite{banica} to quantum mechanics
\cite{znidaric,novaes,berko}).

Previous works where Weingarten functions were obtained were based either on
representation theory and Schur-Weyl duality,\cite{collins,CS} the theory of Gelfand
pairs,\cite{mats} or Jucys-Murphy elements.\cite{zinn} In contrast, we here derive
Weingarten functions for the classical compact groups by means of some elementary direct
calculations (although we rely on some classical results that can, of course, be
interpreted very naturally in the light of those theories).

The idea consists of five steps: 1)
write the integrand as the derivative of a power sum function; 2) change basis from power
sums to Schur functions; 3) perform the group integral; 4) revert back to power sums; 5)
take the derivative to arrive at the result.

\section{Required facts}

Let us review some well known facts and establish notation and terminology. More
detailed explanations of the concepts introduced below can be found in the classical
monograph by MacDonald \cite{donald}.

Everywhere, $\lambda\vdash n$ means $\lambda$ is a partition of $n$, i.e. a weakly
decreasing sequence of positive integers such that $\sum_i \lambda_i=n$. The number of
non-zero parts is called the length of the partition and denoted by $\ell(\lambda)$. The
partition with $n$ unit parts is denoted $1^n$. We define
$2\lambda\equiv(2\lambda_1,2\lambda_2,...)$ and
$\lambda\cup\lambda\equiv(\lambda_1,\lambda_1,\lambda_2,\lambda_2,...)$.

\subsection{Lie groups}

The unitary group $\mathcal{U}(N)$ is the group of complex $N\times N$ matrices satisfying $U^\dag U=1$,
where $U^\dag$ is the transpose conjugate of $U$. 

The orthogonal group $\mathcal{O}(N)$ is the
subgroup of $\mathcal{U}(N)$ of real matrices. This is the isometry group in $\mathbb{R}^N$ with respect to the scalar product
$\{u,v\}=\sum_{i=1}^N u_iv_i$.

Let $J=\left(\begin{array}{cc}0 & 1 \\
-1 & 0\end{array}\right)$, where $0$ and $1$ are, respectively, the zero and the
identity matrix in $N$ dimensions. The unitary symplectic group $Sp(2N)$ is the subgroup of
$\mathcal{U}(2N)$ of complex unitary matrices satisfying $U^DU=1$, where $U^D=JU^TJ^T$. If we define \be\label{prod12}\llangle
u,v\rrangle=\sum_{i=1}^{2N} u_i(J v)_{i}=\sum_{i=1}^{N} u_iv_{i+N}-u_{i+N}v_i,\ee
then $U^D$ plays the role of adjoint of $U$, in the sense that $\llangle
u,Uv\rrangle=\llangle U^Du,v\rrangle$.

\subsection{Permutation groups}

The group of permutations of $n$ elements is $S_n$. The cycle type of $\pi\in S_n$ is the
partition $\lambda\vdash n$ whose parts are the lengths of the cycles of $\pi$ (the cycle
type of the identity in $S_n$ is $1^n$). We denote by $\mathcal{C}_\lambda$ the conjugacy
class of all permutations of cycle type $\lambda$. The length of a permutation is
$\ell(\pi)=\ell(\lambda)$ if $\pi\in\mathcal{C}_\lambda$, and the sign is
$s(\pi)=(-1)^{n-\ell(\pi)}$. We multiply permutations from right to left, e.g.
$(13)(12)=(123)$.

The quantity $\chi_\lambda(\mu)=\chi_\lambda(\pi)$ is the character of the irreducible
representation of $S_n$ labeled by $\lambda$, calculated for a permutation $\pi\in
\mathcal{C}_\mu$. The value $d_\lambda=\chi_\lambda(1^n)$ is the dimension of said
representation. Characters satisfy the orthogonality relation \be \sum_{\tau\in
S_n}\chi_\mu(\tau) \chi_\lambda(\tau\sigma)=
\delta_{\mu,\lambda}n!\frac{\chi_\lambda(\sigma)}{d_\lambda}.\ee

The group $S_{2n}$ has a subgroup called the hyperoctahedral, $H_n$, with $|H_n|=2^nn!$
elements, which is the stabilizer of the permutation $(12)(34)\cdots (2n-1\; 2n)$. The
coset $S_{2n}/H_n=\M_n$ can be represented by permutations called matchings: $\sigma\in
\M_n$ if and only if $\sigma(2i-1)<\sigma(2i)$ and $\sigma(2i-1)<\sigma(2i+1)$.

Given a permutation $\tau\in S_{2n}$, let $\mathcal{G}_\tau$ be a graph with vertices
labeled from $1$ to $2n$ and edges of the forms $\{2i-1,2i\}$ and
$\{\tau(2i-1),\tau(2i)\}$, $1\leq i\leq n$. Since each vertex belongs to one edge of each
form, all connected components of $\mathcal{G}_\tau$ are cycles of even length. The coset
type of $\tau$ is the partition of $n$ whose parts are half the number of edges in the
connected components of $\mathcal{G}_\tau$.

Two permutations $\tau,\sigma$ in $S_{2n}$
have the same coset type if and only if they belong to the same double coset, i.e. if
$\tau=h_1\sigma h_2$ with $h_1,h_2,\in H_n$. We denote by $[\tau]$ the coset type of
$\tau$. We may denote by $H_\tau$ the double coset of $\tau$. Its order is \be |H_\tau|=\frac{4^nn!|\mathcal{C}_{[\tau]}|}{2^{\ell([\tau])}}.\ee

Introduce a map from $S_n$ to $S_{2n}$, taking a permutation
$\pi\in\mathcal{C}_\lambda$ into another permutation
$\widetilde{\pi}\in\mathcal{C}_{2\lambda}$. This map is such that the image of the cycle
$(i_1,i_2,\cdots, i_r)$ in $\pi$ is the cycle $(2i_1-1,2i_1,2i_2-1,2i_2,\cdots,
2i_r-1,2i_r)$ in $\widetilde{\pi}$. This map is useful because the coset type of
$\widetilde{\pi}$ is $\lambda$.

The average \be \omega_\lambda(\tau)=\frac{1}{|H_n|}\sum_{\xi \in
H_n}\chi_{2\lambda}(\tau\xi)\ee is called a zonal spherical function. It is obviously
invariant under (left and right) action of $H_n$ and hence depends only on the coset type
of its argument. These functions satisfy \be\label{omegas} \sum_{\tau\in
\M_{n}}\omega_\lambda(\tau)\omega_\mu(\tau^{-1}\sigma)=\delta_{\lambda,\mu}
\frac{(2n)!}{2^nn!}\frac{\omega_\lambda(\sigma)}{d_{2\lambda}},\ee and also \be
\sum_{\lambda\vdash
n}d_{2\lambda}\omega_\lambda(\sigma)\omega_\lambda(\tau)=\frac{(2n)!}{|H_\tau|}\delta_{[\sigma],[\tau]}.\ee

The average \be \psi_\lambda(\tau)=\frac{1}{|H_n|}\sum_{\xi \in
H_n}\chi_{\lambda\cup\lambda}(\tau\xi)s(\xi),\ee where $s(\xi)$ is the sign of $\xi$, is
called a twisted zonal spherical function. Under action by $H_n$ it is invariant up to
sign: $\psi_\lambda(\tau\xi)=s(\xi)\psi_\lambda(\tau)$. These functions satisfy \be
\sum_{\tau\in \M_{n}}\psi_\lambda(\tau)\psi_\mu(\tau^{-1}\sigma)=\delta_{\lambda,\mu}
\frac{(2n)!}{2^nn!}\frac{\psi_\lambda(\sigma)}{d_{\lambda\cup\lambda}},\ee and also \be
\sum_{\lambda\vdash
n}d_{\lambda\cup\lambda}\psi_\lambda(\sigma)\psi_\lambda(\tau)=\frac{(2n)!}{|H_\sigma|}s(\sigma)s(\tau)\delta_{[\sigma],[\tau]}.\ee

\subsection{Power sum symmetric functions}

Power sum symmetric functions of matrix argument are given by \be
p_\lambda(X)=\prod_{i=1}^{\ell(\lambda)} p_{\lambda_i}(X),\ee with $p_m(X)={\rm
Tr}(X^m)$. They are clearly symmetric functions of the eigenvalues of $X$. If $X$ has
dimension $N$ and $\pi\in S_n$ has cycle type $\lambda$, then\be
p_{\pi}(X)=p_\lambda(X)=\sum_{i_1,...,i_n=1}^N\prod_{k=1}^nX_{i_ki_{\pi(k)}}.\ee

Power sums are produced from some summations. Let $\delta_{ab}$ denote the usual
Kronecker delta function. Given $j=(j_1,j_2,...j_n)$ and $m=(m_1,m_2,...,m_n)$, define
the function \be \delta_\tau[j,m]=\prod_{k=1}^n\delta_{j_km_{\tau(k)}},\ee for any
$\tau\in S_n$. Then we have \be\label{power1} \sum_{j_1\cdots j_n=1}^N\sum_{m_1\cdots
m_n=1}^N \delta_\tau[j,m]\delta_\sigma[j,m]\prod_{k=1}^n
y_{j_k}y_{m_k}=p_{\tau^{-1}\sigma}(x),\ee where \be\label{x2y} x_i=y_i^2,\quad 1\leq
i\leq N.\ee

Let $e(j)$ be the vector with elements given by $(e(j))_i=\delta_{ij}$. With the
definitions from Section 2.1 we have \be \{ e(j_1),e(j_2)\}=\delta_{j_1,j_2},\quad
\llangle e(j_1),e(j_2)\rrangle=\delta_{j_1+N,j_2}-\delta_{j_1,j_2+N}.\ee For any
$\tau\in S_{2n}$, define \be\label{Del}
\Delta_\tau[j]=\prod_{k=1}^n\{e(j_{\tau(2k-1)}),e(j_{\tau(2k)})\},\ee and
\be\label{Delprime} \Delta'_\tau[j]=\prod_{k=1}^n \llangle
e(j_{\tau(2k-1)}),e(j_{\tau(2k)})\rrangle. \ee Notice that the behavior of these
functions under action of the hyperoctahedral group is $\Delta_{\tau\xi}=\Delta_\tau$ and
$\Delta'_{\tau\xi}=s(\xi)\Delta'_\tau$. The summations that produce power sums are
\be\label{power2} \sum_{j_1\cdots j_{2n}=1}^N
\Delta_{\tau}[j]\Delta_{\sigma}[j]\prod_{k=1}^{2n} y_{j_k}=p_{[\tau^{-1}\sigma]}(x),\ee
and \be\label{power3} \sum_{j_1\cdots j_{2n}=1}^{2N}
\Delta'_{\tau}[j]\Delta'_{\sigma}[j]\prod_{k=1}^{2n}
y_{j_k}=(-1)^ns(\tau)s(\sigma)(-2)^{\ell([\tau^{-1}\sigma])}p_{[\tau^{-1}\sigma]}(x),\ee where $[\tau]$ is the coset
type of $\tau$, while the variables satisfy $y_{i+N}=y_i$ and (\ref{x2y}).

\subsection{Jack polynomials}

Power sums and Schur functions are related by \be s_\lambda(X)=\sum_{\pi\in
S_n}\chi_\lambda(\pi)p_\pi(X)=\sum_{\mu\vdash
n}|\mathcal{C}_\mu|\chi_\lambda(\mu)p_\mu(X), \ee and \be p_\lambda(X)=\sum_{\mu\vdash
n}\chi_\mu(\lambda)s_\mu(X).\ee Schur functions are generalized by Jack polynomials,
$J_\lambda^{(\alpha)}(x)$, which depend on the parameter $\alpha$. They are recovered when
$\alpha=1$: \be J_\lambda^{(1)}(x)=\frac{n!}{d_\lambda}s_\lambda(x).\ee The polynomials
$J_\lambda^{(2)}$ and $J_\lambda^{(1/2)}$ are called zonal polynomials and quaternion
zonal polynomials, respectively. The former are related to power sums according to
\be\label{JtopO} J_\lambda^{(2)}(X)=\sum_{\mu\vdash n}
|\mathcal{C}_\mu|2^{n-\ell(\mu)}\omega_\lambda(\mu)p_\mu(X),\ee and \be
p_{[\tau]}(x)=\frac{2^nn!}{(2n)!}\sum_{\lambda\vdash n}d_{2\lambda}
\omega_\lambda(\tau)J_\lambda^{(2)}(x).\ee For the latter, the first relation is
\be\label{JtopS} J_\lambda^{(1/2)}(X)=(-1)^{n}\sum_{\mu\vdash
n}|\mathcal{C}_\mu|\psi_\lambda(\widetilde{\pi})p_\mu(X),\ee where $\pi$ is any
permutation of cycle type $\mu$ and $\widetilde{\pi}$ is its image under the map
introduced in Section 2.2. The second relation is \be p_{[\tau]}(x)=\frac{(-4)^nn!}{(2n)!}
\left(\frac{-1}{2}\right)^{\ell([\tau])}\sum_{\lambda\vdash n}d_{\lambda\cup\lambda}
s(\tau)\psi_\lambda(\tau)J_\lambda^{(1/2)}(x),\ee valid for any $\tau\in S_{2n}$.

The value of the Jack polynomial when all arguments are equal to $1$ is given by
\be\label{J1N} J_\lambda^{(\alpha)}(1^N)=\alpha^n\prod_{j=1}^{\ell(\lambda)}
\frac{\Gamma(\lambda_j+(N-j+1)/\alpha)}{\Gamma((N-j+1)/\alpha)}.\ee 

Jack polynomials arise from integrals of Schur functions over Lie groups. Let $dU$, $dO$
and $dS$ denote the normalized Haar measures over $\mathcal{U}(N)$, $\mathcal{O}(N)$ and
$Sp(2N)$, respectively. The orthogonality relation \be\label{orthog} \int_{\mathcal{U}(N)}
dU s_\lambda(A^\dag U)s_\mu(BU^\dag)=\delta_{\mu,\lambda}\frac{J^{(1)}_\lambda(A^\dag
B)}{J^{(1)}_\lambda(1^N)}\ee  holds because Schur functions are actually irreducible
characters of the unitary group. We also have the relations \be\label{relO}
\int_{\mathcal{O}(N)}dO s_\mu(AO)=\delta_{\mu,2\lambda}\frac{J^{(2)}_\lambda(A^T
A)}{J^{(2)}_\lambda(1^N)},\ee and \be\label{averageS}\int_{Sp(2N)}dS
s_\mu(AS)=\delta_{\mu,\lambda\cup\lambda}2^{-\ell(\lambda)}\frac{J^{(1/2)}_\lambda(A^DA)}{J^{(1/2)}_\lambda(1^N)}.\ee
In the last equation $A$ and $S$ are $2N\times 2N$ complex matrices. All the above integrals vanish
unless $\ell(\lambda)\leq N$.

\section{Derivation}

\subsection{Unitary Group}

We start from the basic identity \be \prod_{k=1}^n
U_{a_kb_k}=\frac{1}{n!}\prod_{k=1}^n\frac{\partial }{\partial A^*_{a_kb_k}}
p_{1^n}(A^\dag U),\ee which can be easily verified. We express power sums in terms of
Schur functions, use (\ref{orthog}) to perform the group integral, and then revert back
to power sums to get \be \int_{\mathcal{U}(N)} dU \prod_{k=1}^n
U_{a_kb_k}U^*_{d_kc_k}=\frac{1}{n!^2}\sum_{\substack{\lambda\vdash n\\\ell(\lambda)\leq
N}} \frac{d_\lambda}{J^{(1)}_\lambda(1^N)}\sum_{\pi \in
S_n}\chi_\lambda(\pi)\prod_{k=1}^n\frac{\partial }{\partial A^*_{a_kb_k}}\frac{\partial
}{\partial B_{m_kr_k}}p_{\pi}(A^\dag B).\ee In order to take the derivative, we expand
$p_\pi$ as a trace, \be p_{\pi}(A^\dag
B)=\sum_{i_1,...,i_n=1}^N\sum_{j_1,...,j_n=1}^NA^\dag_{i_kj_k}B_{j_ki_{\pi(k)}}.\ee
Taking the derivative and summing over the $i$'s and $j$'s produces \be
\sum_{\theta,\rho\in
S_n}\prod_{k=1}^n\delta_{a_k,d_{\rho\theta^{-1}(k)}}\delta_{b_k,c_{\rho\pi^{-1}\theta^{-1}(k)}}.\ee

Finally, we change variables as $\rho=\sigma\theta$ and then $\tau=\sigma\theta
\pi^{-1}\theta^{-1}$. This last change preserves cycle type, so that
$\chi_\lambda(\pi)=\chi_\lambda(\sigma^{-1}\tau)=\chi_\lambda(\tau^{-1}\sigma)$. The sum
over $\pi$ is then just $n!$, and we arrive at \be  \int_{\mathcal{U}(N)} dU
\prod_{k=1}^n U_{a_kb_k}U^*_{d_kc_k}=\sum_{\sigma\tau \in S_n}{\rm
Wg}^U(\tau^{-1}\sigma)\delta_\sigma[ad]\delta_\tau[bc],\ee where \be {\rm
Wg}^U(\tau^{-1}\sigma)=\frac{1}{n!}\sum_{\substack{\lambda\vdash n\\\ell(\lambda)\leq N}}
\frac{d_\lambda}{J^{(1)}_\lambda(1^N)}\chi_\lambda(\tau^{-1}\sigma),\ee is the unitary
Weingarten function.

\subsection{Orthogonal Group}

In the same spirit, we start by writing \be \prod_{k=1}^{2n}
O_{a_kb_k}=\frac{1}{(2n)!}\prod_{k=1}^{2n}\frac{\partial }{\partial A_{a_kb_k}}
p_{1^{2n}}(A^T O).\ee Changing to Schur functions and using (\ref{relO}) we get
\be\frac{1}{(2n)!}\sum_{\substack{\lambda\vdash n\\\ell(\lambda)\leq N}}
\frac{d_{2\lambda}}{J^{(2)}_\lambda(1^N)}\prod_{k=1}^{2n}\frac{\partial }{\partial
A_{a_kb_k}}J^{(2)}_\lambda(A^T A).\ee We return to power sums in order to take the
derivative. Taking $\pi\in S_n$ to be any permutation with cycle type $\mu$, we write \be
\int_{\mathcal{O}(N)} dO\prod_{k=1}^{2n}
O_{a_kb_k}=\frac{1}{(2n)!}\sum_{\substack{\lambda\vdash n\\\ell(\lambda)\leq N}}
\frac{d_{2\lambda}}{J^{(2)}_\lambda(1^N)}\sum_{\mu\vdash
n}|\mathcal{C}_\mu|2^{n-\ell(\mu)}\omega_\lambda(\mu)\prod_{k=1}^{2n}\frac{\partial
}{\partial A_{a_kb_k}}p_\pi(A^T A).\ee

The derivative \be \prod_{k=1}^{2n}\frac{\partial }{\partial A_{a_kb_k}}p_\pi(A^TA)=
\sum_{i_1,...,i_n=1}^N\sum_{j_1,...,j_n=1}^N\prod_{r=1}^{n}\frac{\partial }{\partial
A_{a_{(2r-1)}b_{(2r-1)}}}\frac{\partial }{\partial
A_{a_{(2r)}b_{(2r)}}}A^T_{i_{r}j_{r}}A_{j_{r}i_{\pi(r)}},\ee will, upon summing over the
$i$'s and $j$'s, always lead to a matching among the $a$'s and a matching among the
$b$'s, i.e. something like $\Delta_\tau[a]\Delta_\sigma[b]$. By construction, these
matchings will satisfy $[\tau^{-1}\sigma]=\mu$, and each such pair will appear exactly
$2^{\ell(\mu)}n!/|\mathcal{C}_\mu|$ times for any $\pi$. Hence, we can write
\be\int_{\mathcal{O}(N)} dO\prod_{k=1}^{2n} O_{a_kb_k}=\sum_{\tau,\sigma\in \M_n}{\rm
Wg}^O(\tau^{-1}\sigma)\Delta_\tau(a)\Delta_\sigma(b),\ee where \be {\rm
Wg}^O(\tau^{-1}\sigma)=\frac{2^nn!}{(2n)!}\sum_{\substack{\lambda\vdash
n\\\ell(\lambda)\leq N}}
\frac{d_{2\lambda}}{J^{(2)}_\lambda(1^N)}\omega_\lambda(\tau^{-1}\sigma)\ee is the
Weingarten function of $\mathcal{O}(N)$.

\subsection{Symplectic Group}

Let $\pi\in S_n$ be any permutation of cycle type $\mu$ and $\widetilde{\pi}$ be its
image under the map in Section 2.2, so that $[\widetilde{\pi}]=\mu$. Calculations
analogous to those in the previous section lead to \be\int_{Sp(2N)} dS\prod_{k=1}^{2n}
S_{a_kb_k}=\frac{(-1)^n}{(2n)!}\sum_{\substack{\lambda\vdash n\\\ell(\lambda)\leq N}}
\frac{d_{\lambda\cup\lambda}}{J^{(1/2)}_\lambda(1^N)}\sum_{\mu\vdash
n}|\mathcal{C}_\mu|2^{-\ell(\mu)}\psi_\lambda(\widetilde{\pi})\prod_{k=1}^{2n}\frac{\partial
}{\partial A_{a_kb_k}}p_\pi(A^D A),\ee where we have used (\ref{JtopS}) and
(\ref{averageS}). The derivative $\prod_{k=1}^{2n}\frac{\partial }{\partial
A_{a_kb_k}}p_\pi(A^DA)$ leads to a matching among the $a$'s and a matching among the
$b$'s, but with such minus signs as to produce  $(-1)^n\Delta'_\tau[a]\Delta'_\sigma[b]$.
Again, these matchings satisfy $[\tau^{-1}\sigma]=\mu$, and each such pair appears
$2^{\ell(\mu)}n!/|\mathcal{C}_\mu|$ times for any $\pi$. Therefore, \be \int_{Sp(2N)}
dS\prod_{k=1}^{2n} S_{a_kb_k}=\sum_{\tau,\sigma\in \M_n}{\rm
Wg}^{Sp}(\tau^{-1}\sigma)\Delta'_\tau[a]\Delta'_\sigma[b],\ee where \be {\rm
Wg}^{Sp}(\tau^{-1}\sigma)=\frac{n!}{(2n)!}\sum_{\substack{\lambda\vdash
n\\\ell(\lambda)\leq N}}
\frac{d_{\lambda\cup\lambda}}{J^{(1/2)}_\lambda(1^N)}\psi_\lambda(\tau^{-1}\sigma)\ee is
the Weingarten function of $Sp(2N)$.

\section*{Acknowledgments} Financial support from CNPq is gratefully acknowledged, as well as from grant 2012/00699-1, S˜\~ao Paulo
Research Foundation (FAPESP).

\end{document}